\documentclass[pre,twocolumn,aps]{revtex4}
\newcommand{\br}{{\bf r}}
\newcommand{\ba}{{\bf a}}
\newcommand{\bb}{{\bf b}}
\newcommand{\bR}{{\bf R}}
\newcommand{\bS}{{\bf S}}
\newcommand{\bD}{{\bf\Delta}}
\newcommand{\bF}{{\bf F}}

\newcommand{\rmu}{{\rm U}}
\usepackage{graphicx}
\usepackage{epstopdf}
\usepackage{bm}
\begin{document}
\title{$NVU$ dynamics. I. Geodesic motion on the constant-potential-energy hypersurface}
\author{Trond S. Ingebrigtsen, S{\o}ren Toxvaerd, Ole J. Heilmann, Thomas B. Schr{\o}der, and Jeppe C. Dyre}
\email{dyre@ruc.dk}
\affiliation{DNRF Centre ``Glass and Time'', IMFUFA, Department of Sciences, Roskilde University, Postbox 260, DK-4000 Roskilde, Denmark}
\date{\today}

\begin{abstract}
An algorithm is derived for computer simulation of geodesics on the constant potential-energy hypersurface of a system of $N$ classical particles. First, a basic time-reversible geodesic algorithm is derived by discretizing the geodesic stationarity condition and implementing the constant potential energy constraint via standard Lagrangian multipliers. The basic $NVU$ algorithm is tested by single-precision computer simulations of the Lennard-Jones liquid. Excellent numerical stability is obtained if the force cutoff is smoothed and the two initial configurations have identical potential energy within machine precision. Nevertheless, just as for $NVE$ algorithms, stabilizers are needed for very long runs in order to compensate for the accumulation of numerical errors that eventually lead to ``entropic drift'' of the potential energy towards higher values. A modification of the basic $NVU$ algorithm is introduced that ensures potential-energy and step-length conservation; center-of-mass drift is also eliminated. Analytical arguments confirmed by simulations demonstrate that the modified $NVU$ algorithm is absolutely stable. Finally, simulations show that the $NVU$ algorithm and the standard leap-frog $NVE$ algorithm have identical radial distribution functions for the Lennard-Jones liquid. 
\end{abstract}

\maketitle

\section{Introduction}\label{intr}

This paper and its companion Paper II \cite{II} study $NVU$ dynamics, i.e., dynamics that conserves the potential energy $U$ for a system of $N$ classical particles at constant volume $V$. $NVU$ dynamics is deterministic and involves only the system's configurational degrees of freedom. $NVU$ dynamics is characterized by the system moving along a so-called {\it geodesic} curve on the constant potential-energy hypersurface $\Omega$ defined by 

\begin{equation}\label{omega_def}
\Omega\,=\,
\{(\br_1,...,\br_N)\in R^{3N}|U(\br_1,...,\br_N)=U_0\}\,.
\end{equation}
Mathematically, $\Omega$ is a $3N-1$ dimensional differentiable manifold. Since it is imbedded in $R^{3N}$, $\Omega$ has a natural Euclidean metric and it is thus a so-called Riemannian manifold \cite{diff_geom}. The differential geometry of hypersurfaces is discussed in, for instance, Ref. \onlinecite{hyp}. 

A geodesic curve on a Riemannian manifold by definition minimizes the distance between any two of its points that are sufficiently close to each other (the curve is characterized by realizing the ``locally shortest distance'' between points). More generally, a geodesic is defined by the property that for any curve variation keeping the two end points $\bR_A$ and $\bR_B$ fixed, to lowest order the curve length does not change, i.e.,

\begin{equation}\label{geod}
  \delta \int_{\bR_A}^{\bR_B} dl \,=\, 0\,.
\end{equation}
Here $dl$ denotes the line element of the metric. 

From a physical point of view it is sometimes useful to regard a geodesic as a curve along which the system moves at constant velocity with zero friction. Such motion means that at any time the force is perpendicular to the surface, and because the force performs no work, the kinetic energy is conserved. In this way geodesic motion generalizes Newton's first law, the law of inertia, to curved surfaces. The concept of geodesic motion is central  in general relativity, where motion in a gravitational field follows a geodesic curve in the four-dimensional curved space-time \cite{gen_rel}.

A general motivation for studying $NVU$ dynamics is the following. Since all relevant information about a system is encoded in the potential-energy function, it is interesting from a philosophical point of view to study and compare different dynamics relating to $U(\br_1,...,\br_N)$. The ``purest'' of these dynamics does not involve momenta and relates only to configuration space. $NVU$ dynamics provides such a dynamics. In contrast to Brownian dynamics, which also relates exclusively to the configurational degrees of freedom, $NVU$ dynamics is deterministic. $NVU$ dynamics may be viewed as an attempt to understand the dynamic implications of the potential energy landscape's geometry along the lines of recent works by Stratt and coworkers \cite{wan07,ngu10}.

Our interest in $NVU$ dynamics originated in recent results concerning strongly correlating liquids and their isomorphs. A liquid is termed strongly correlating if there is more than 90\% correlation between its virial and potential energy thermal equilibrium fluctuations in the $NVT$ ensemble \cite{scl}. The class of strongly correlating liquids includes most or all van der Waals and metallic liquids, whereas hydrogen-bonding, covalently bonded liquids, and ionic liquids are generally not strongly correlating.  A liquid is strongly correlating if and only if it to a good approximation has ``isomorphs'' in its phase diagram \cite{IV,Rome}. By definition two state points are isomorphic \cite{IV} if any two microconfigurations of the state points, which can be trivially scaled into one another, have identical canonical probabilities; an isomorph is a curve in the phase diagram for which any two pairs of points are isomorphic. Only inverse-power-law liquids have exact isomorphs, but simulations show that Lennard-Jones type liquids have isomorphs to a good approximation \cite{IV}. This is consistent with these liquids being strongly correlating \cite{scl}. Many properties are invariant along an isomorph, for instance the excess entropy, the isochoric heat capacity, scaled radial distribution functions, dynamic properties in reduced units, etc \cite{IV,Rome}; the reduced-unit constant-potential-energy hypersurface $\tilde\Omega$ is also invariant along an isomorph \cite{IV}. Given that several properties are invariant along a strongly correlating liquid's isomorphs and that $\tilde\Omega$ is invariant as well, an obvious idea is that $\tilde\Omega$'s invariance is the fundamental fact from which all other isomorph invariants follow. For instance, the excess entropy is the logarithm of the area of $\tilde\Omega$, so the excess entropy's isomorph invariance follows directly from that of $\tilde\Omega$. In order to understand the dynamic isomorph invariants from the $\tilde\Omega$ perspective a dynamics is required that refers exclusively to $\tilde\Omega$. One possibility is diffusive dynamics, but a mathematically even more elegant dynamics on a differentiable manifold is that of geodesics. Although these considerations were our original motivation, it should be emphasized that that the concept of geodesic motion on $\tilde\Omega$ (or $\Omega$) is general and makes sense for any classical mechanical system, strongly correlating or not.

We are not the first to consider dynamics on the constant-potential-energy hypersurface. In papers dating back to 1986 \cite{cott} Cotterill and Madsen proposed a deterministic constant-potential-energy algorithm similar, but not identical, to the basic $NVU$ algorithm derived below. Their algorithm was not discussed in relation to geodesic curves, but aimed at providing an alternative way to understanding vacancy diffusion in crystals and, in particular, to make easier the identification of energy barriers than from ordinary MD simulations. The latter property is not confirmed in the present papers, however -- we find that $NVU$ dynamics in the thermodynamic limit becomes equivalent to standard $NVE$ dynamics (Paper II \cite{II}). Later Scala {\it et al.} studied diffusive dynamics on the constant-potential-energy hypersurface $\Omega$ \cite{sca02}, focusing on the entropic nature of barriers by regarding these as ``bottlenecks''. This point was also made by Cotterill and Madsen who viewed $\Omega$ as consisting of  ``pockets'' connected by thin paths, referred to as ``tubes'', acting as entropy barriers. Reasoning along similar lines, Stratt and coworkers published in 2007 and 2010 three papers \cite{wan07,ngu10}, which considered paths in the so-called potential-energy-landscape ensemble. This novel ensemble is defined as including all configurations with potential energy less than or equal to some potential energy $U_0$. A geodesic in the potential-energy-landscape ensemble consists of a curve that is partly geodesic on the constant-potential-energy surface $\Omega$, partly a straight line in the space defined by  $U<U_0$ \cite{wan07}. Stratt {\it et al.}'s picture shifts ``perspective from finding stationary points on the potential energy landscape to finding and characterizing the accessible pathways through the landscape. Within this perspective pathways would be slow, not because they have to climb over high barriers, but because they have to take a long and tortuous route to avoid such barriers...." \cite{wan07}. Thus the more ``convoluted and laborinthine'' the geodesics are, the slower is the dynamics \cite{wan07}. Apart from these three sources of inspiration to the present work, we note that geodesic motion on differentiable manifolds has been studied in several other contexts outside of pure mathematics, see, e.g., Ref. \onlinecite{div_const_U}. 

The present paper derives and documents an algorithm for $NVU$ geodesic dynamics. In Sec. II we derive the basic $NVU$ algorithm. By construction this algorithm is time reversible, a feature that ensures a number of important properties \cite{shaddow,constraint}. Section III discusses how to implement the $NVU$ algorithm and tests improvements of the basic $NVU$ algorithm designed for ensuring stability, which is done by single-precision simulations. This section arrives at the final $NVU$ algorithm and demonstrates that it conserves potential energy, step length, and center-of-mass position in arbitrarily long simulations. Section IV briefly investigates the sampling properties of the $NVU$ algorithm, showing that it gives results for the Lennard-Jones liquid that are equivalent to those of standard $NVE$ dynamics. Finally, Sec. V gives some concluding comments. Paper II compares $NVU$ simulations to results for four other dynamics, concluding that $NVU$ dynamics is a fully valid molecular dynamics.

\section{The basic $NVU$ algorithm}

For simplicity of notation we consider in this paper only systems of particles of identical masses (Appendix A of Paper II generalizes the algorithm to systems of varying particle masses). The full set of positions in the $3N$-dimensional configuration space is collectively denoted by $\bR$, i.e., 

\begin{equation}\label{R}
\bR\,\equiv\,(\br_1,...,\br_N)\,.
\end{equation}
Likewise, the full $3N$-dimensional force vector is denoted by $\bF$. This section derives the basic $NVU$ algorithm for geodesic motion on the constant-potential-energy hypersurface $\Omega$ defined in Eq. (\ref{omega_def}), an algorithm that allows one to compute the positions in step $i+1$, $\bR_{i+1}$, from $\bR_{i-1}$ and $\bR_i$. Although a mathematical geodesic on a differentiable manifold is usually parameterized by its curve length \cite{diff_geom}, it is useful to think of a geodesic curve on $\Omega$ as parameterized by time, and we shall refer to the steps of the algorithm as ``time steps''.

Locally, a geodesic is the shortest path between any two of its points. More precisely: 1) For any two points on a Riemannian manifold the shortest path between them is a geodesic; 2) The property of a curve being geodesic is locally defined; 3) a geodesic curve has the property that for any two of its points, which are sufficiently close to each other, the curve gives the shortest path between them. A geodesic may, in fact, be the {\it longest} distance between two of its points. For instance, the shortest and the longest flight between two cities on our globe both follow great circles -- these are both geodesics. In any case, the property of being geodesic is always equivalent to the curve length being {\it stationary} in the following sense: Small curve variations, which do not move the curve's end points, to lowest order lead to no change in the curve length. 

For motion on $\Omega$ the constraint of constant potential energy is taken into account by introducing Lagrangian multipliers. For each time step $j$ there is the constraint $U(\bR_j)=U_0$ and a corresponding Lagrangian multiplier $\lambda_j$. Thus the stationarity condition Eq. (\ref{geod}) for the discretized curve length $\sum_j|\bR_j-\bR_{j-1}|$ subject to the constraint of constant potential energy, is

\begin{equation}\label{nvu2}
\delta \left(\sum_j|\bR_j-\bR_{j-1}|-\sum_j\lambda_j U( \bR_j)\right)\,=\,0\,.
\end{equation}
Since $|\bR_j-\bR_{j-1}|=\sqrt{(\bR_j-\bR_{j-1})^2}$ and the $3N$-dimensional force is given by $\bF_j=-\partial U/\partial \bR_j$, putting to zero the variation of Eq. (\ref{nvu2}) with respect to $\bR_i$ (i.e., the partial derivative $\partial/\partial \bR_i$) leads to

\begin{equation}\label{nvu3}
\frac{\bR_i-\bR_{i-1}}{|\bR_i-\bR_{i-1}|} -\frac{\bR_{i+1}-\bR_{i}}{|\bR_{i+1}-\bR_{i}|} + \lambda_i \bF_i\,=\,0\,.
\end{equation}
To solve these equations we make the ansatz of constant displacement length for each time step,

\begin{equation}\label{nvu4}
|\bR_j-\bR_{j-1}|\equiv l_0\,\,\, ({\rm all}\,\,\,j)\,.
\end{equation}
If the path discretization is thought of as defined by constant time increments, Eq. (\ref{nvu4}) corresponds to constant velocity in the geodesic motion. With this ansatz Eq. (\ref{nvu3}) becomes

\begin{equation}\label{nvu5}
(\bR_{i}-\bR_{i-1})+(\bR_{i}-\bR_{i+1}) +l_0\lambda_i \bF_i\,=\,0\,.
\end{equation}
If $\ba_i\equiv \bR_{i}-\bR_{i-1}$ and $\bb_i\equiv \bR_{i}-\bR_{i+1}$, Eq. (\ref{nvu4}) implies $\ba_i^2=\bb_i^2$, i.e., $0=\ba_i^2-\bb_i^2=(\ba_i+\bb_i)\cdot(\ba_i-\bb_i)$. Since Eq. (\ref{nvu5}) expresses that $\ba_i+\bb_i$ is parallel to $\bF_i$, one concludes that $\bF_i$ is perpendicular to $\ba_i-\bb_i=\bR_{i+1}-\bR_{i-1}$. This implies

\begin{equation}\label{nvu6}
\bF_i\cdot \bR_{i-1}\,=\,\bF_i\cdot \bR_{i+1}\,.
\end{equation}
Taking the dot product of each side of Eq. (\ref{nvu5}) with $\bF_i$ one gets

\begin{equation}\label{nvu7}
\bF_i\cdot(\bR_{i}-\bR_{i-1})+\bF_i\cdot(\bR_{i}-\bR_{i+1})+l_0\lambda_i \bF_i^2\,=\, 0\, ,
\end{equation}
which via Eq. (\ref{nvu6}) implies

\begin{equation}\label{nvu_lagr}
l_0\lambda_i\,=\, -2\,\frac{\bF_i\cdot(\bR_i-\bR_{i-1})}{ \bF_i^2}\,.
\end{equation}
Substituting this into Eq. (\ref{nvu5}) and isolating $\bR_{i+1}$ we finally arrive at 

\begin{equation}\label{nvu1}
\bR_{i+1}\,=\,
2\bR_i-\bR_{i-1}-2 [\bF_i\cdot\left( \bR_i-\bR_{i-1}\right)]\bF_i/\bF_i^2\,.
\end{equation}
This equation determines a sequence of positions; it will be referred to as ``the basic $NVU$ algorithm''. The algorithm is  initialized by choosing two nearby points in configuration space with the same potential energy within machine precision. 

The derivation of the basic $NVU$ algorithm is completed by checking its consistency with the constant step length ansatz Eq. (\ref{nvu4}): Rewriting Eq. (\ref {nvu1}) as $(\bR_{i+1}-\bR_i)=(\bR_i-\bR_{i-1}) -2 [\bF_i\cdot\left( \bR_i-\bR_{i-1}\right)]\bF_i/\bF_i^2$ we get by squaring each side $(\bR_{i+1}-\bR_i)^2=(\bR_i-\bR_{i-1})^2 +4[\bF_i\cdot\left( \bR_i-\bR_{i-1}\right)]^2/\bF_i^2- 4[\bF_i\cdot\left( \bR_i-\bR_{i-1}\right)]^2/\bF_i^2=(\bR_i-\bR_{i-1})^2 $. Thus the solution is consistent with the ansatz.

Time reversibility of the basic $NVU$ algorithm is checked by rewriting Eq. (\ref{nvu1}) as follows

\begin{equation}\label{11}
\bR_{i-1}\,=\,
2\bR_i-\bR_{i+1}-2 [\bF_i\cdot\left( \bR_i-\bR_{i-1}\right)]\bF_i/\bF_i^2\,,
\end{equation}
which via Eq. (\ref{nvu6}) implies 

\begin{equation}\label{12}
\bR_{i-1}\,=\,
2\bR_i-\bR_{i+1}-2 [\bF_i\cdot\left( \bR_i-\bR_{i+1}\right)]\bF_i/\bF_i^2\,.
\end{equation}
Comparing to Eq. (\ref{nvu1}) shows that any sequence of configurations generated by Eq. (\ref{nvu1}) $...,\bR_{i-1}, \bR_i, \bR_{i+1},...$ obeys Eq. (\ref{nvu1}) in the time-reversed version $..., \bR_{i+1}, \bR_i, \bR_{i-1}, ...$. A more physical way to show that the basic $NVU$ algorithm is time-reversal invariant is to note that Eq. (\ref{nvu3}) is itself manifestly invariant if the indices $i-1$ and $i+1$ are interchanged. 

Appendix A shows that the basic $NVU$ algorithm is symplectic, i.e., that it conserves the configuration-space volume element in the same way as $NVE$ dynamics does. We finally consider potential-energy conservation in the basic $NVU$ algorithm. A Taylor expansion implies via Eq. (\ref{nvu6}) that 

\begin{equation}\label{13}
U_{i+1}-U_{i-1}\,=\,
-\bF_i\cdot \left( \bR_{i+1}-\bR_{i-1}\right)+O(l_0^3)\,=\,O(l_0^3)\,.
\end{equation}
This ensures potential-energy conservation to a good approximation if the discretization step is sufficiently small.

The ``potential energy contour tracing'' (PECT) algorithm of Cotterill and Madsen \cite{cott} is the following: $\bR_{i+1}=2\bR_i-\bR_{i-1}- [\bF_i\cdot\left( \bR_i-\bR_{i-1}\right)]\bF_i/\bF_i^2$. Except for a factor of $2$ this is identical to the basic $NVU$ algorithm. The importance of this difference is apparent when it is realized that the PECT algorithm implies $\bF_i\cdot (\bR_{i+1}-\bR_i)=0$, whereas it does not imply the time-reversed identity $\bF_i\cdot (\bR_{i-1}-\bR_i)=0$. Thus the PECT algorithm is not time reversible.

We end this section by reflecting on what is the relation between the $NVU$ algorithm and continuous geodesic curves on $\Omega$. Can one expect that if the step length is decreased towards zero, the discrete sequence of points traced out by the algorithm converges to a continuous geodesic curve? The answer is yes, as is clear from the current literature that treats this problem in considerable detail \cite{disc_var}. The literature deals with the analogous problem of classical mechanics where, as is well known, Newton's second law of motion can be derived from the principle of least action (Hamilton's principle). This is a variational principle. In the traditional approach one  first derives continuous equations of motion from the variational principle, then discretizes these equations to allow for computer simulations. Here we first discretized the quantity subject to the variational principle (Eq. (\ref{nvu2})) and only thereafter applied variational calculus. 

Euler himself first described discretization of time in the action integral, thus obtaining discretized versions of the Euler-Lagrange equations. There is now a large literature on this subject \cite{disc_var}. During the last decade, in particular, variational calculations applied {\it after} discretization have come into focus in connection with for instance the development of algorithms for the control of robots. The general consensus is the following (we quote below from Ref. \onlinecite{ste08} that provides an excellent, brief  summary of the situation): ``The driving idea behind this discrete geometric mechanics is to leverage the variational nature of mechanics and to preserve this variational structure in the discrete setting... If one designs a discrete equivalent of the Lagrangian, then discrete equations of motion can be easily derived from it by paralleling the derivations followed in continuous case, and good numerical methods will come from discrete analogs to the Euler-Lagrange equations. In essence, good numerical methods will come from discrete analogs to the Euler-Lagrange equations -- equations that truly derive from a variational principle...Results have been shown to be equal or superior to all other types of integrators for simulations of a large range of physical phenomenon, making this discrete geometric framework both versatile and powerful.''

\section{Testing and improving the basic $NVU$ algorithm}

This section discusses the numerical implementation of the basic $NVU$ algorithm and how to deal with round-off errors that arise for very long simulations. The model system studied is the standard Lennard-Jones (LJ) liquid with  $N=1024$ particles. Recall that the LJ pair potential $v(r)$ is given by

\begin{equation}\label{lj_pot}
v(r)\,=\,
 4\varepsilon \left[\left(\frac{\sigma}{r}\right)^{12}-\left(\frac{\sigma}{r}\right)^{6}\right]\,.
\end{equation}
Here $ \varepsilon$ sets the energy scale and $\sigma$ the length scale; henceforth the unit system is adopted in which these quantities are both unity. All simulations except those of Fig. \ref{sampling} refer to the state point with density 0.85 and temperature 0.7 in reduced units. The two initial configurations were taken from $NVE$ simulations of this state point. Unless otherwise specified the forces and their derivative were adjusted to be continuous via smoothing from a value  just below the cutoff distance $r_c$ to $r_c$. We refer to this as a ``smoothed force potential''. The cutoff distance was chosen as the standard LJ cutoff $r_c = 2.5 \sigma$. The simulations were performed using periodic boundary conditions. In order to easier test the numerical stability of the $NVU$ algorithm, simulations were performed in single precision \cite{simulations}.

\subsection{Implementing the basic $NVU$ algorithm}

We rewrite Eq. (\ref{nvu1}) into a leap-frog version by introducing new variables defined by

\begin{equation}\label{bD_def}
\bD_{i+1/2}\,\equiv\,\bR_{i+1}-\bR_i\,.
\end{equation}
In terms of these variables the basic $NVU$ algorithm is 

\begin{eqnarray}\label{nvu_lp}
\bD_{i+1/2}\,&=&\,\bD_{i-1/2}-2(\bF_i\cdot\bD_{i-1/2})\bF_i/\bF_i^2\nonumber\\
\bR_{i+1}\,&=&\,\bR_i+\bD_{i+1/2}\,.
\end{eqnarray}
The equations (\ref{nvu_lp}) are formally equivalent to Eq. (\ref{nvu1}). Numerically, however, they are not equivalent and -- as is also the case for standard $NVE$ dynamics -- the leap-frog version is preferable because it deals with position changes \cite{sim}.

\begin{figure}
  \centering
  \includegraphics[width=70mm]{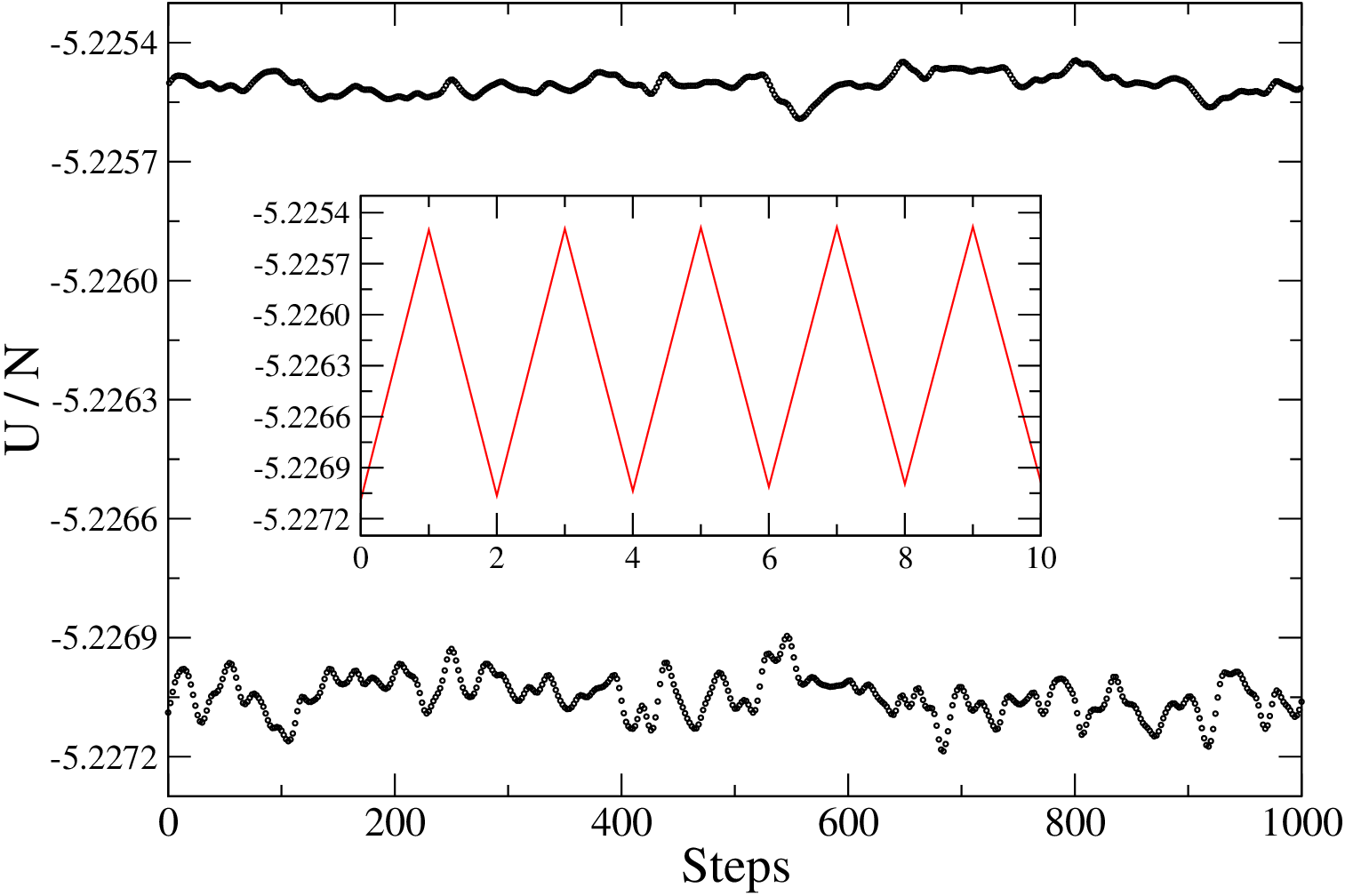}
  \includegraphics[width=70mm]{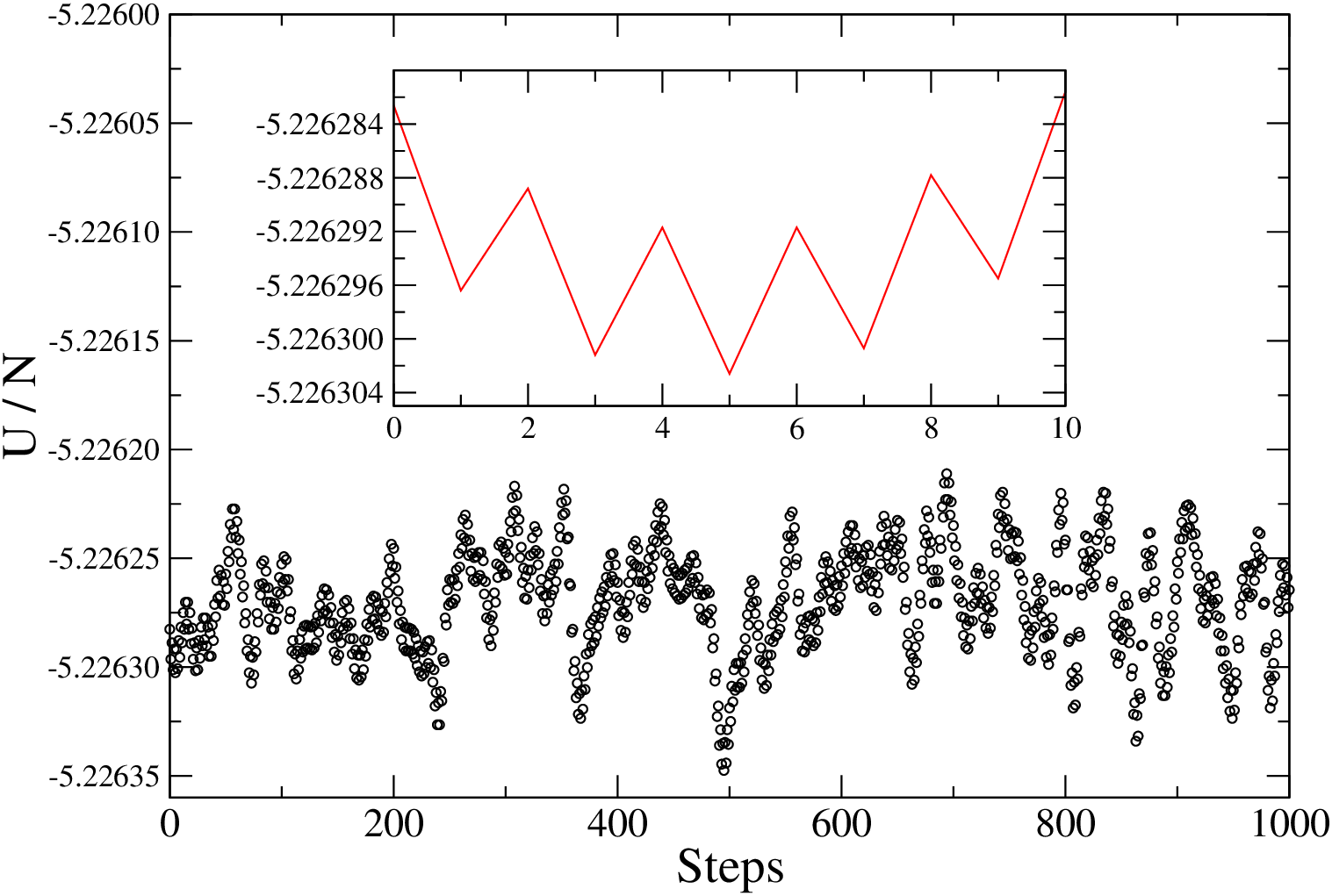}
  \includegraphics[width=70mm]{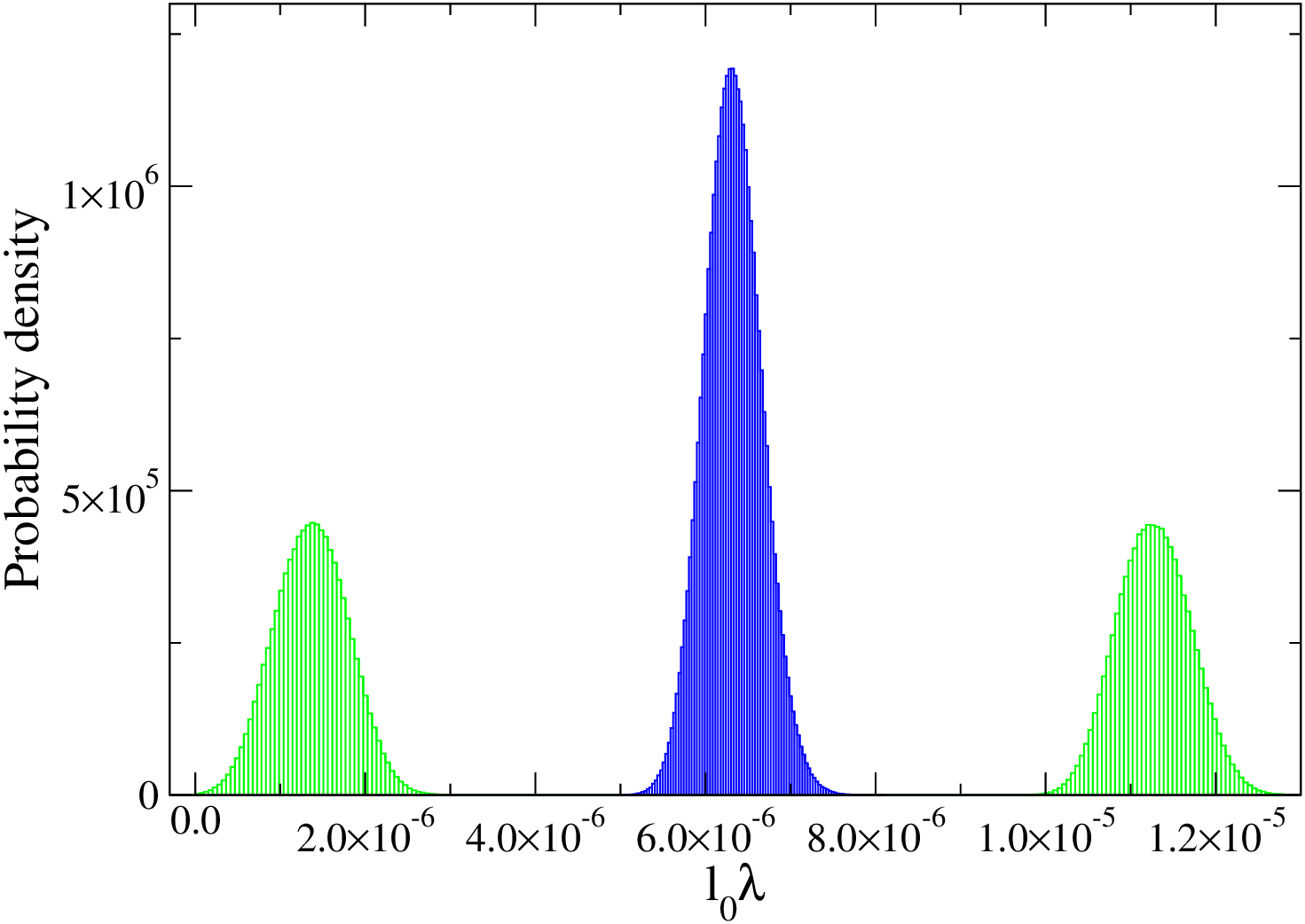}
  \caption{\label{U_figs} 
(a) Evolution of the potential energy $U$ according to the basic $NVU$ algorithm (Eq. (\ref{nvu_lp})) started from two consecutive configurations of an $NVE$ simulation. The inset shows a snapshot of the first ten integration steps where lines connect the data points; clearly the system jumps distinctly between two potential-energy hypersurfaces. 
(b)  Evolution of $U$ started from two configurations with a very small potential energy difference. The algorithm still jumps between two potential-energy hypersurfaces, but the difference is much smaller.
(c) Probability distribution of the Lagrangian multiplier times the length $l_0$, $l_0\lambda$ of Eq. (\ref{nvu_lagr}), obtained from simulations over 2.5 $\cdot 10^{6}$ steps. The green distribution corresponds to (a), the blue distribution to (b).}
\end{figure}

Figure \ref{U_figs}(a) shows the potential energy as a function of time-step number. The system's potential energy jumps every second step, jumping between two distinct values (inset). This is also reflected in the distribution of the quantity $l_0\lambda_i$ shown in green in Fig. \ref{U_figs}(c). {\it A priori} one would expect a Gaussian single-peak distribution of $l_0\lambda_i$, but the distribution has two peaks. What causes the potential energy to zig-zag in an algorithm constructed to conserve potential energy? The answer is hinted at in Eq. (\ref{13}) according to which the $NVU$ algorithm implies energy conservation to a good accuracy, but only every second step. Thus if the two initial configurations do not have identical potential energy, the potential energy will zig-zag between two values. Figure \ref{U_figs}(b) shows that even if a simulation is initiated from two configurations with very close potential energies, the zig-zag phenomenon persists, though now on a much smaller scale.

\begin{figure}
  \centering
  \includegraphics[width=70mm]{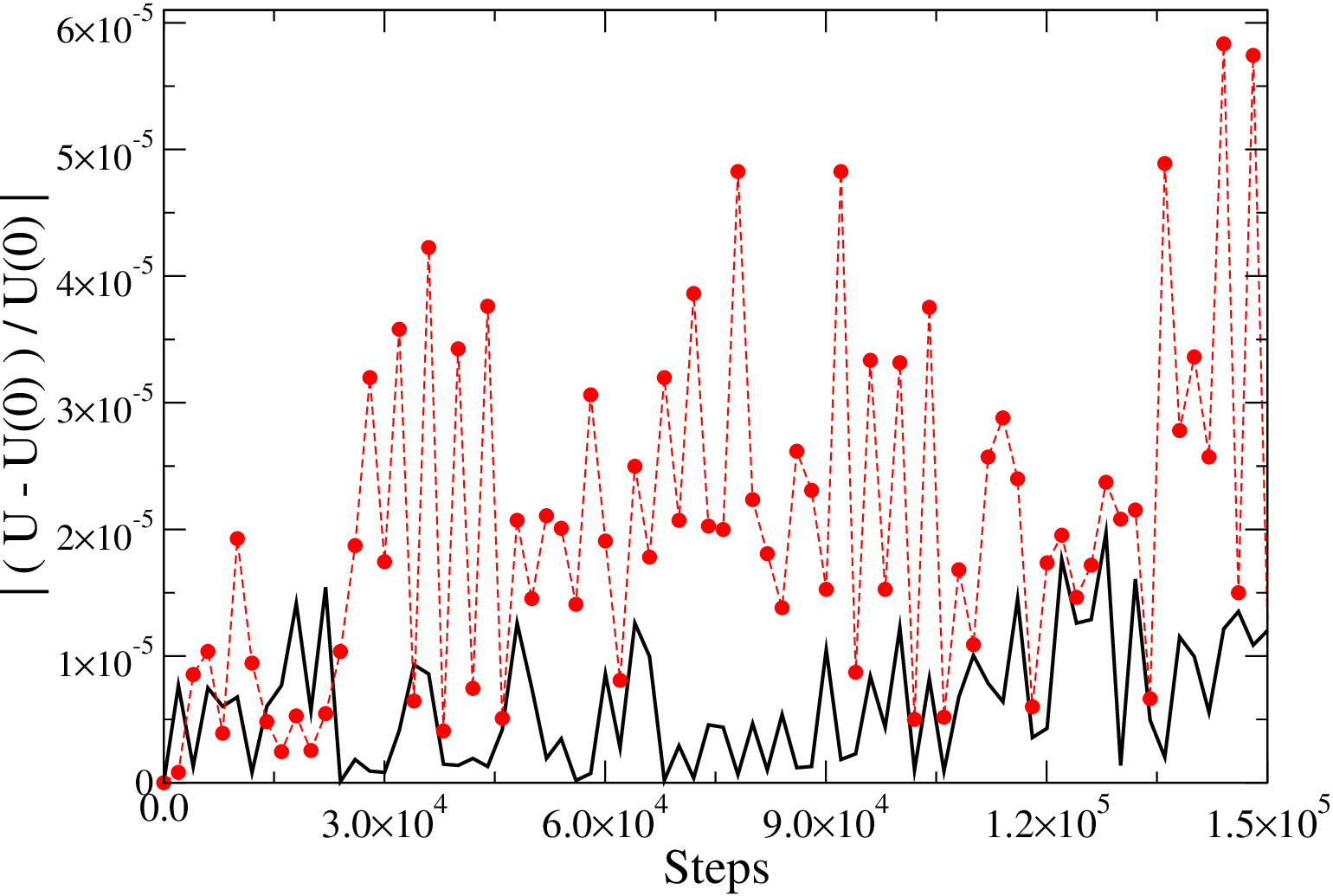}
 \caption{\label{smoothe}
Evolution of $|(U - U(0)) / U(0)|$ for a simulation using the basic $NVU$ algorithm. The red curve gives results from a simulation where the potential is cut and shifted at $r=2.5\sigma$, the black curve gives results for a smoothed force potential.
}
\end{figure}

There are further numerical issues that effect the stability of the basic $NVU$ algorithm. In Fig. \ref{smoothe} the evolution of the potential energy is given for a long simulation, which also includes data from simulations using a smoothed force potential. Better numerical stability is clearly obtained for the smoothed force potential (black curve), but smoothing does not ensure a constant potential energy and absolute stability.

\subsection{Improving the algorithm to conserve potential energy and step length indefinitely}

\begin{figure}
  \centering
  \includegraphics[width=70mm]{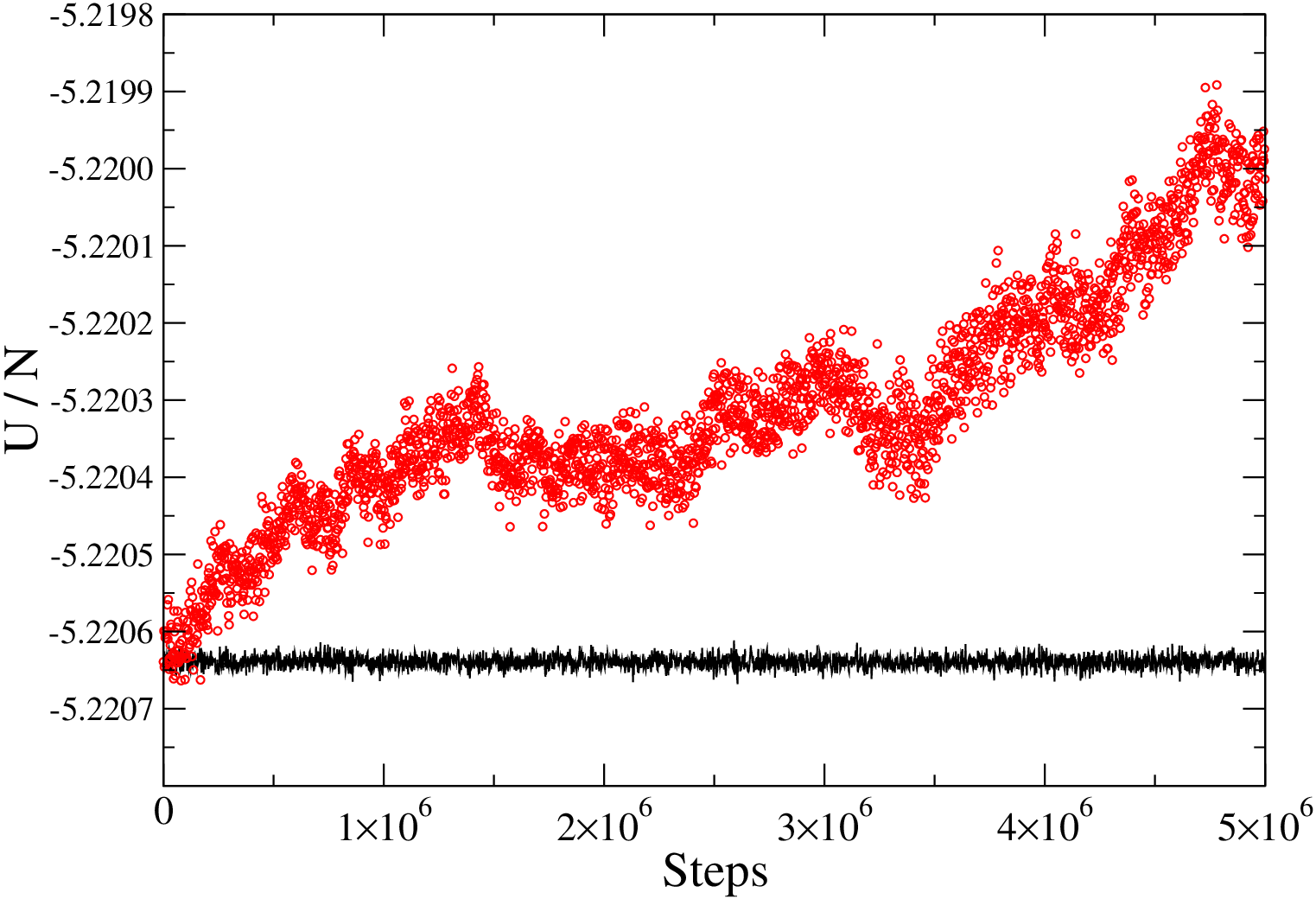}
  \includegraphics[width=70mm]{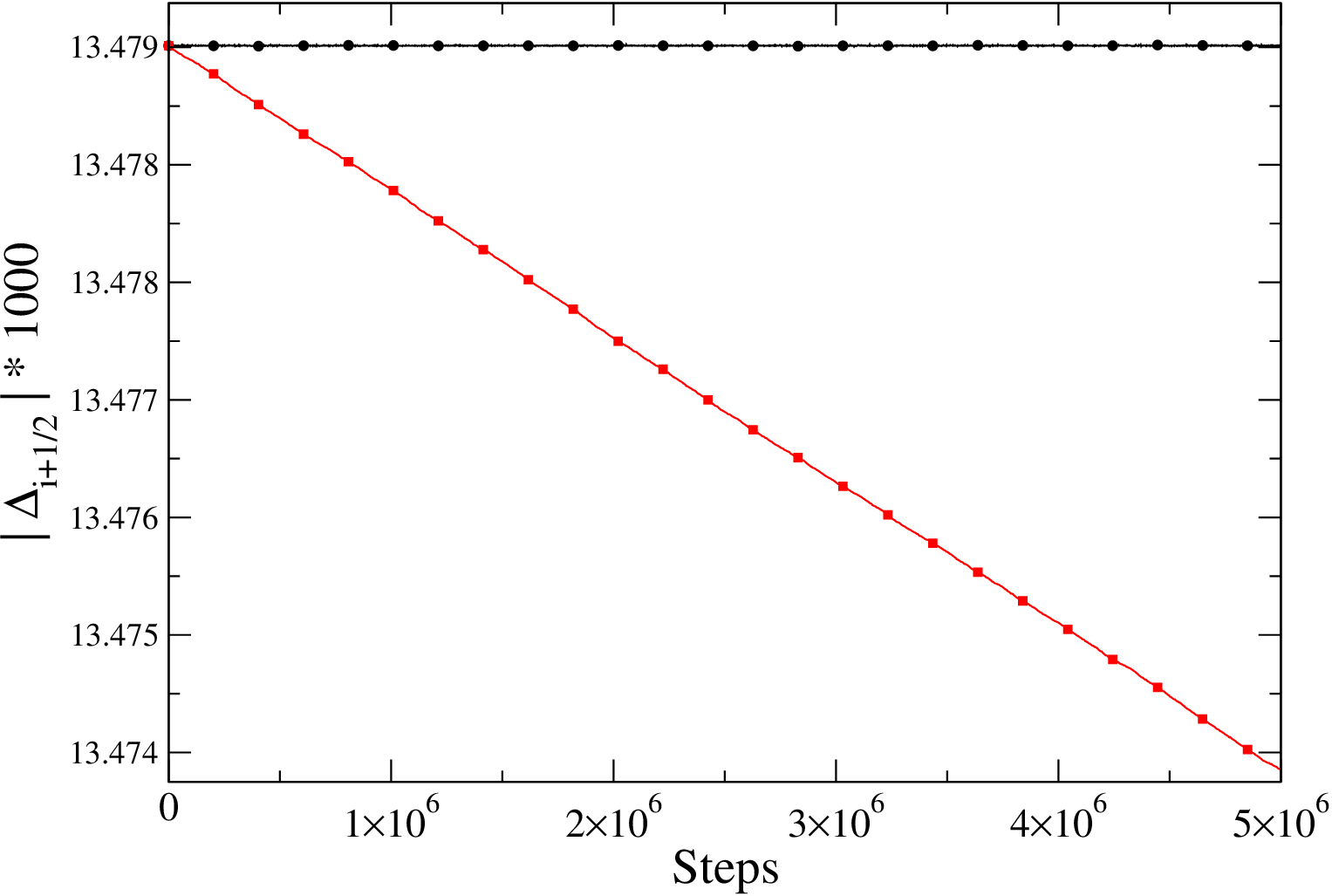}
  \caption{\label{drift_new}
(a) Evolution of $U$ with and without the numerical stabilization (Eq. (\ref{tt})): The red curve gives results using the basic $NVU$ algorithm Eq. (\ref{nvu_lp}) with two identical initial potential energies and smoothed force potential. The black curve gives simulation results under the same conditions using the final $NVU$ algorithm (Eq. (\ref{tt})).
(b) Evolution of the step length $|\bD_{i}|$ for the same simulations.}
\end{figure}

The last subsection showed that using a smoothed force potential and ensuring that the two starting configurations have identical potential energy within machine precision, a more stable algorithm is arrived at. Nevertheless, absolute stability is not obtained. This is illustrated in Fig. \ref{drift_new}(a), which shows that the potential energy for a system with a smoothed force potential over five million time steps still exhibits a slight ``entropic drift'' (red curve). By entropic drift we mean the drift due to round-off errors -- a drift that unavoidably takes the system to higher energies because there are many more such states, an entropic effect. Figure \ref{drift_new}(b) shows that also the step length is not conserved. Both problems are caused by the accumulation of round-off errors. These problems are less severe if one switches to double precision, of course, but for long simulations entropic drift eventually sets in (for billions of time steps). 

We would like to have an algorithm that is absolutely stable, i.e., one that does not allow for any long-time drift of quantities the basic $NVU$ algorithm was constructed to conserve: the potential energy, the step length, and the center of mass (CM) position (just as in standard $NVE$ dynamics the CM position is exactly conserved in the basic $NVU$ algorithm Eq. (\ref{nvu1}) because the forces sum to zero due to the translational invariance of the potential energy: $U(\br_1+\br^0, ..., \br_N+\br^0)=U(\br_1, ..., \br_N)$). 

Drift of the CM position is trivially eliminated by adjusting the particle displacements according to $\Delta\br_n =\Delta\br_n - \sum_n \Delta\br_n / N$, every 100'th time steps. This correction corresponds to setting to zero the total momentum of the system in an $NVE$ simulation.  

Robust potential energy conservation is obtained by adding a term that is zero if the potential energy equals the target potential energy $\rmu$ (this quantity was previously denoted by $U_0$, but to avoid confusion with the time step index we drop the subscript zero),

\begin{equation}\label{nvu_trond}
\bD_{i+1/2}\,=\,\bD_{i-1/2}+\Big(-2\bF_i\cdot\bD_{i-1/2}+U_{i-1}-\rmu\Big){\bF_i}/{\bF_i^2}\,.
\end{equation}
To show that this modification of the $NVU$ algorithm prevents drift of the potential energy, we take the dot product of each side of Eq. (\ref{nvu_trond}) with $\bF_i$, leading to $\bF_i\cdot \bD_{i+1/2} = -\bF_i\cdot \bD_{i-1/2} +U_{i-1}-\rmu$ or $\bF_i\cdot (\bD_{i+1/2} +\bD_{i-1/2})=U_{i-1}-\rmu$. Since $\bF_i\cdot (\bD_{i+1/2} +\bD_{i-1/2})=\bF_i\cdot (\bR_{i+1} -\bR_{i-1})=-(U_{i+1}-U_{i-1})+O(l_0^3)$, this implies

\begin{equation}\label{U_conserve}
U_{i+1}\,=\,\rmu+O(l_0^3)\,.
\end{equation}
Thus entropic drift has been eliminated and the potential energy is conserved indefinitely except for small fluctuations.

We next address the problem of conserving step length. This is ensured by the following modification of the algorithm,

\begin{equation}\label{tt}
  \bD_{i+1/2} =  l_0\,\frac{\bD_{i-1/2}+(-2\bF_i\cdot\bD_{i-1/2}+U_{i-1}-\rmu){\bF_i}/{\bF_i^2}}
{\left|\bD_{i-1/2}+(-2\bF_i\cdot\bD_{i-1/2}+U_{i-1}-\rmu){\bF_i}/{\bF_i^2}\right|}\,.
\end{equation}
Equation (\ref{tt}) gives what we term the final $NVU$ algorithm (for brevity: ``the $NVU$ algorithm'', in contrast to Eq. (\ref{nvu1}) that is referred to as ``the basic $NVU$ algorithm''). 

In simulations the $NVU$ algorithm is implemented as follows. The target potential energy $\rmu$ is chosen from an $NVE$ or an $NVT$ simulation at the relevant state point. The step length $l_0$ is chosen according to the accuracy aimed for. Suppose at a given time the quantities $\bR_i$, $\bD_{i-1/2}$, and $U_{i-1}$ are given. From $\bR_i$ the forces $\bF_i$ are calculated. From $\bD_{i-1/2}$, $\bF_i$, and $U_{i-1}$ the quantity $\bD_{i+1/2}$ is calculated via Eq. (\ref{tt}). Finally, the positions are updated via $\bR_{i+1}=\bR_i+\bD_{i+1/2}$ and the potential energy is updated via $U_i=U(\bR_i)$.

By construction the $NVU$ algorithm Eq. (\ref{tt}) ensures constant step length,

\begin{equation}\label{cond}
  |\bD_{i+1/2}|\, =\, l_0\,,
\end{equation}
but is the potential energy still conserved for arbitrarily long runs? If the denominator of Eq. (\ref{tt}) is denoted by $D_i$, taking the dot product of each side of this equation with $\bF_i$ leads to
$\bF_i\cdot \bD_{i+1/2}= (l_0/D_i)\left[-\bF_i\cdot \bD_{i-1/2}+U_{i-1} -\rmu\right]$. Writing $l_0/D_i\equiv 1+\delta_i$ in which $\delta_i=O(l_0^p)$ with $p\ge 1$, we get $\bF_i\cdot (\bD_{i+1/2}+\bD_{i-1/2})= \delta_i\left[-\bF_i\cdot \bD_{i-1/2}\right]+(1+\delta_i)\left[U_{i-1}-\rmu\right]$. Thus, since $\bF_i\cdot (\bD_{i+1/2}+\bD_{i-1/2})=U_{i-1}-U_{i+1}+O(l_0^3)$ and $\bF_i\cdot \bD_{i-1/2}=U_{i-1}-U_{i}+O(l_0^2)$, we get $\rmu-U_{i+1}+O(l_0^3)=\delta_i\left[U_i-\rmu+O(l_0^2)\right]$. This implies again

\begin{equation}\label{delta}
U_{i+1}\,=\,\rmu+O(l_0^3)\,.
\end{equation}

\begin{figure}
  \centering
   \includegraphics[width=70mm]{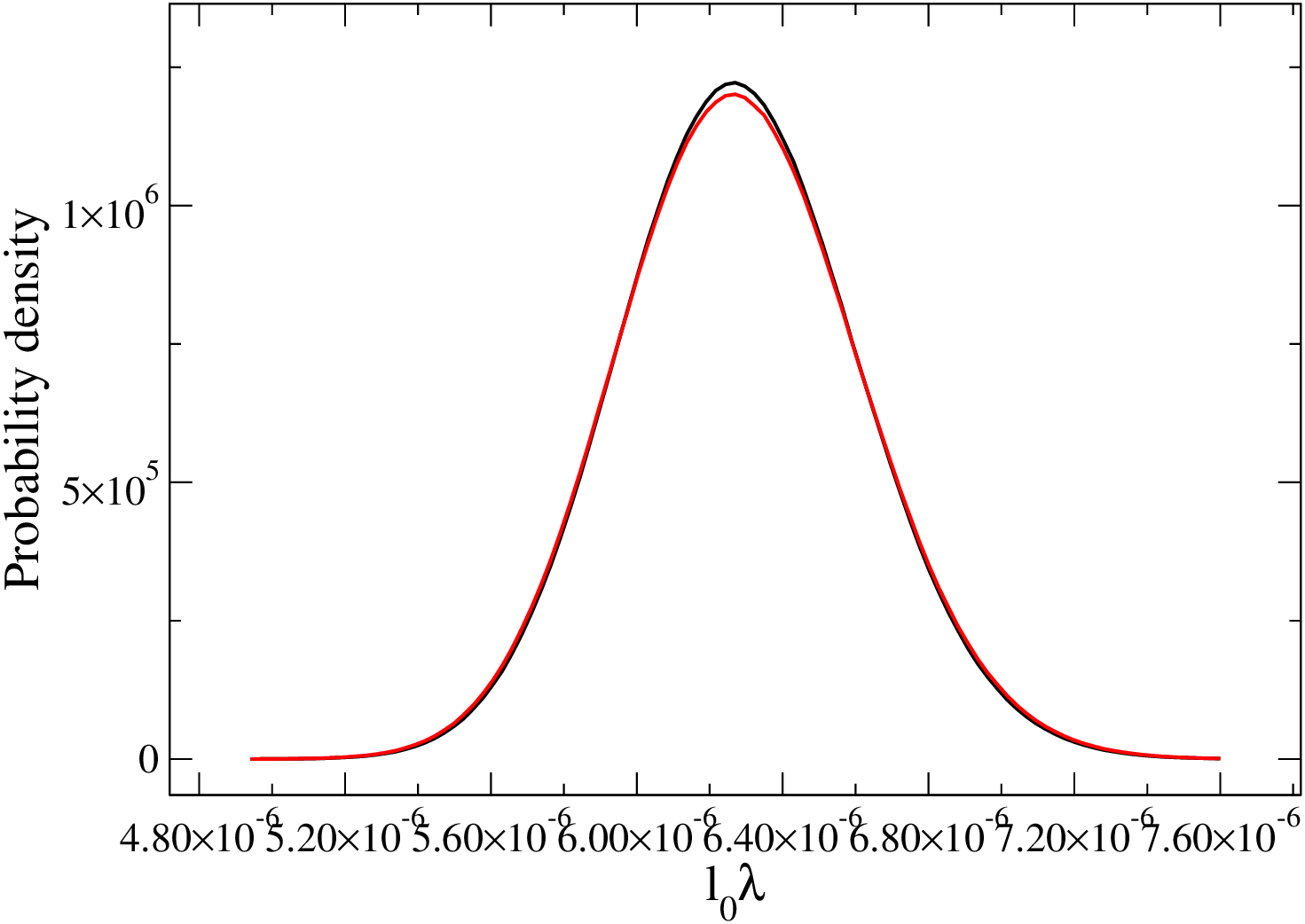}
   \includegraphics[width=70mm]{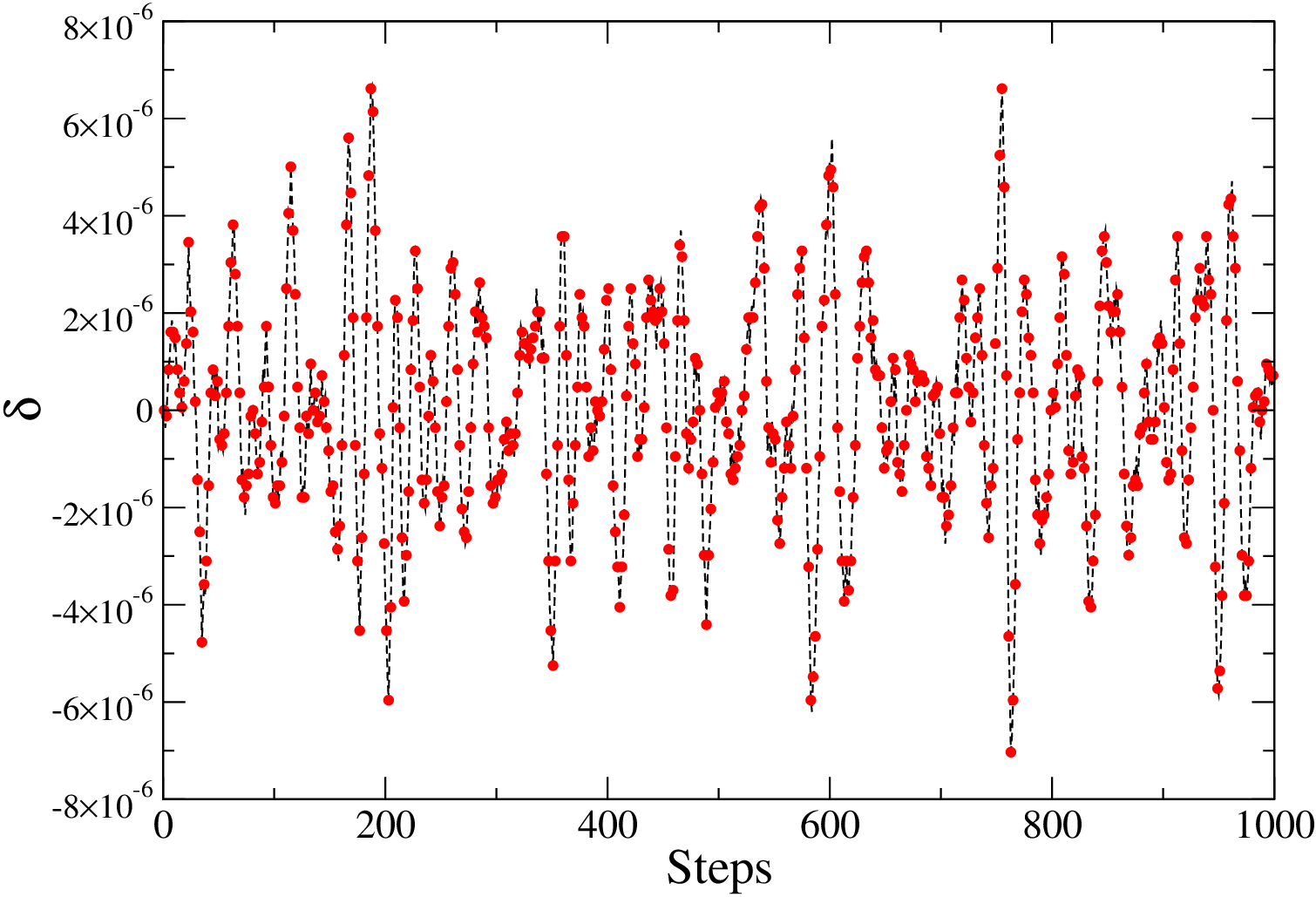}
   \caption{\label{corr}
(a) The distribution of the Lagrangian multiplier times $l_0$ with (black) and without (red) the numerical stabilization of the final $NVU$ algorithm Eq. (\ref{tt}).
(b) Evolution of the quantity $\delta_i$ defined by $l_0/D_i\equiv 1+\delta_i$; as expected this quantity is small and averages to zero.}
\end{figure}

In summary, for simulations of indefinite length the $NVU$ algorithm Eq. (\ref{tt}) ensures constant step length and avoids entropic drift of the potential energy. Figure \ref{drift_new}(a) shows the evolution of the potential energy using the basic $NVU$ algorithm (red) and the final $NVU$ algorithm (black), Fig. \ref{drift_new}(b) shows the analogous step length evolution. Figure \ref{corr}(a) shows that the distribution of the Lagrangian multiplier is only slightly affected by going from the basic (red) to the final (black) $NVU$ algorithm. Figure \ref{corr}(b) shows the evolution of $\delta_i$ in the final $NVU$ algorithm, which as expected is close to zero.

We remind the reader that the modifications were introduced to compensate for the effects of accumulating random numerical errors for very long runs, and that the modifications introduced in the final $NVU$ algorithm Eq. (\ref{tt}) vanish numerically in the mean. The prize paid for stabilizing the basic $NVU$ algorithm is that the full $NVU$ algorithm is not time reversible. In view of the fact that the improvements introduced to ensure stability lead to very small corrections, the (regrettable) fact that the corrections violate time reversibility is not important.

\section{Sampling properties of the $NVU$ algorithm}

\begin{figure}
  \centering
\includegraphics[width=70mm]{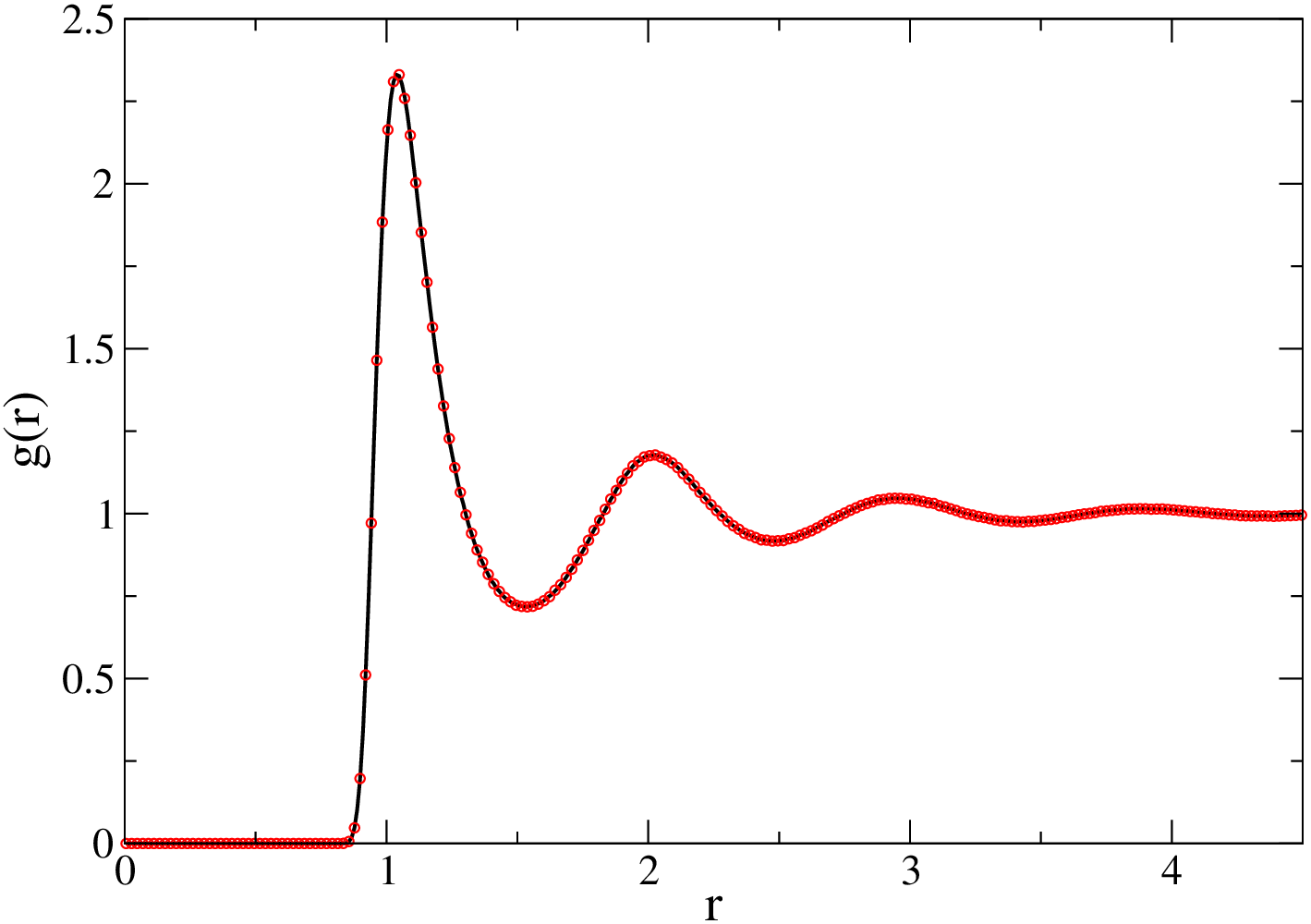}
\includegraphics[width=70mm]{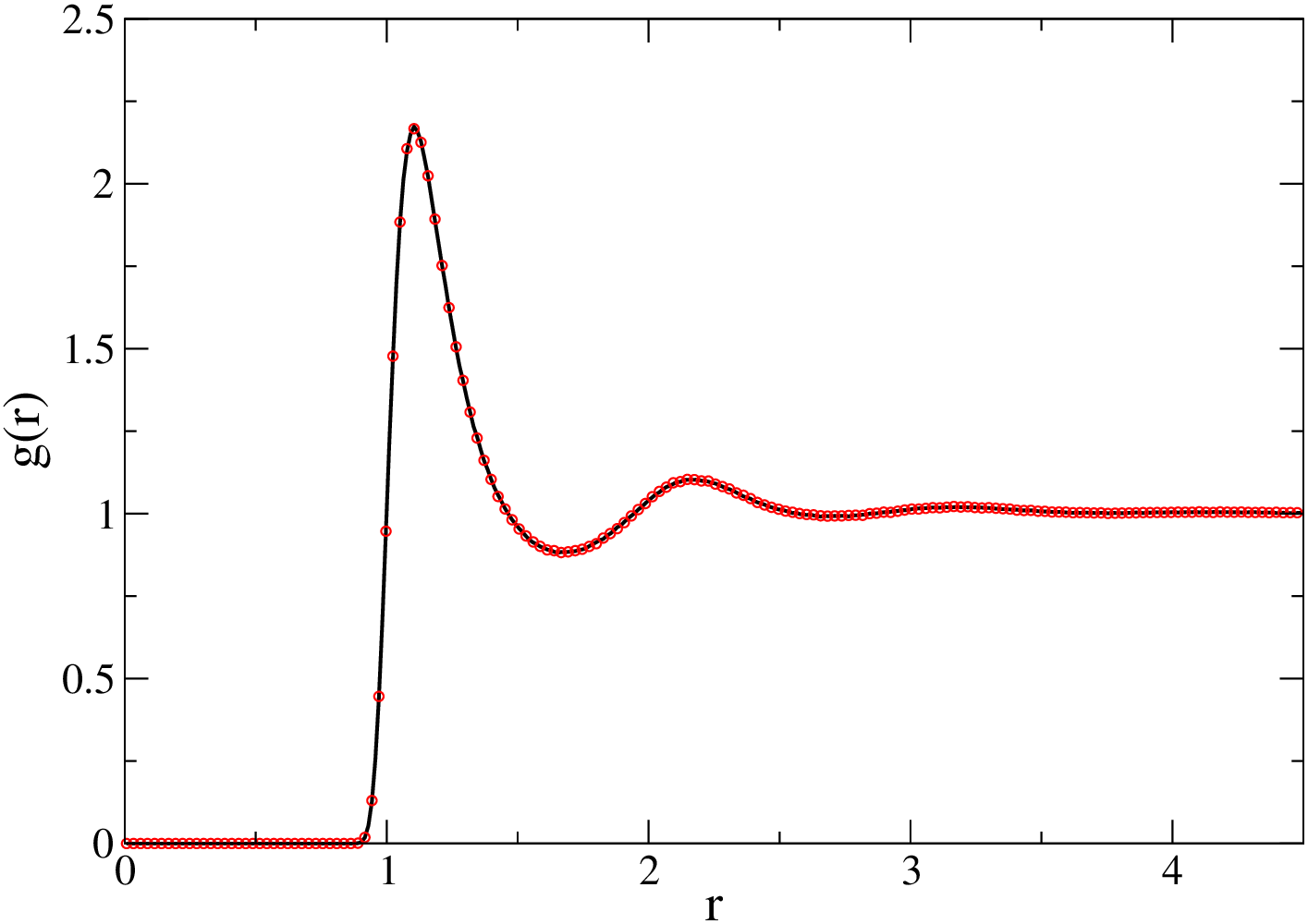}
\includegraphics[width=70mm]{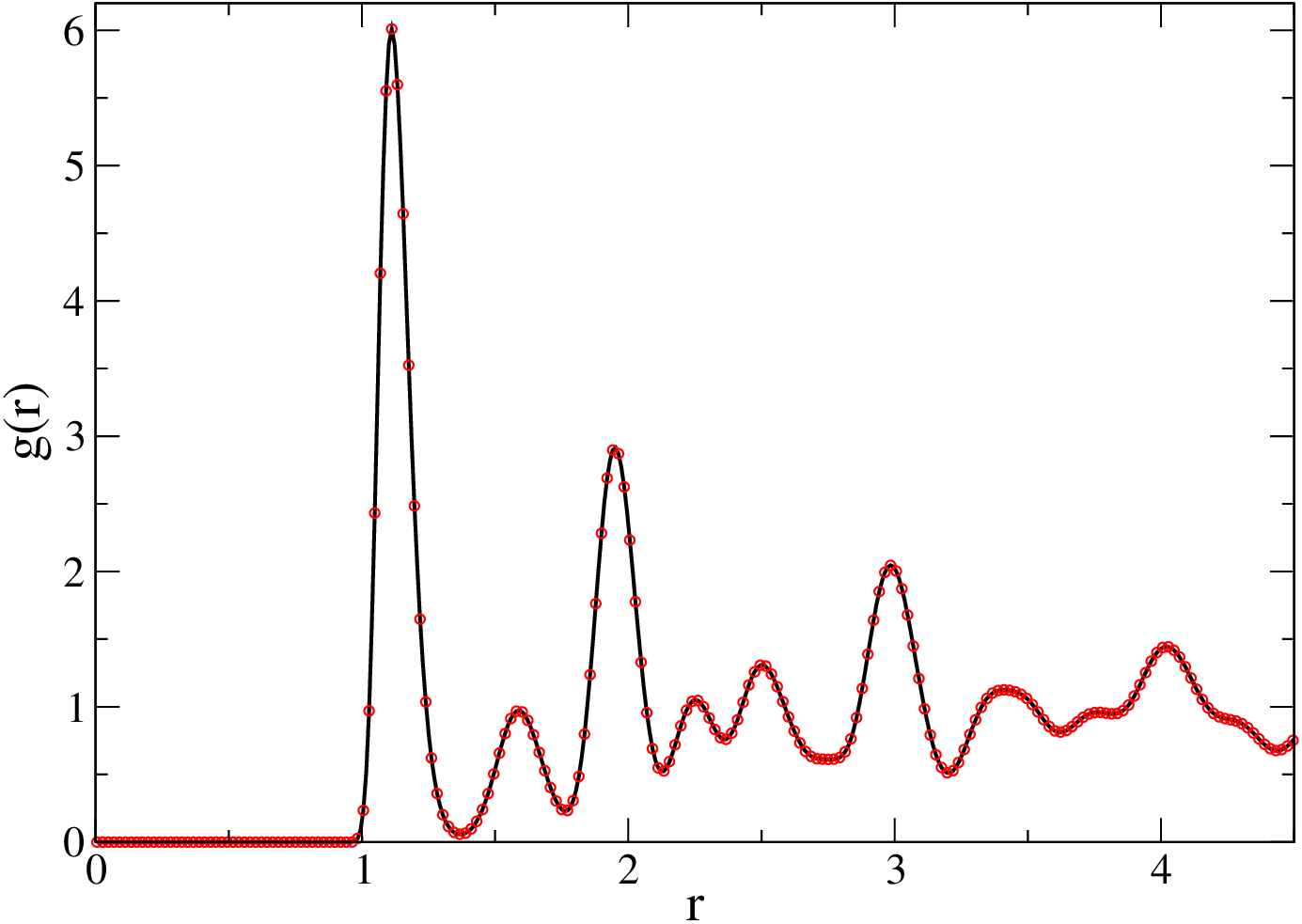}
 \caption{\label{sampling}
Radial distribution functions $g(r)$ for a single component Lennard-Jones system at the following state points: 
(a) $T = 2.32$ and $\rho = 0.85$; 
(b) $T = 1.1$  and $\rho = 0.427$; 
(c) the crystal at $T = 0.28$ and $\rho = 0.85$. 
The black curves show results from $NVE$ simulations, the red dots from $NVU$ simulations (Eq. (\ref{tt})).}
\end{figure}

In order to investigate whether the $NVU$ algorithm gives physically reasonable results we compare results from $NVU$ and $NVE$ simulations for the average of a quantity that depends only on configurational degrees of freedom. This is done in Fig. \ref{sampling}, which shows the radial distribution function $g(r)$ at three state points. The red dots give $NVU$ simulation results, the black curve $NVE$. Clearly the two algorithms give the same results. This finding is consistent with the conjecture that the $NVU$ algorithm probes all points on $\Omega$ with equal probability. Note that this is not mathematically equivalent to conjecturing that the $NVU$ algorithm probes the configuration space microcanonical ensemble, which has equal probability density everywhere in a thin energy shell between a pair of constant-potential-energy manifolds. The latter distribution would imply a density of points on $\Omega$ proportional to the length of the gradient of $U(\bR)$ (the force), but this distribution cannot be the correct equilibrium distribution because the basic $NVU$ algorithm Eq. (\ref{nvu1}) is invariant to local scaling of the force. In the thermodynamic limit, however, the length of the force vector becomes almost constant and the difference between the configuration space microcanonical ensemble and the $\Omega$ equal-measure ensemble becomes insignificant.

Paper II details a comparison of $NVU$ dynamics to four other dynamics, including two stochastic dynamics. Here both simulation and theory lead to the conclusion that $NVU$ and $NVE$ dynamics are equivalent in the thermodynamic limit.

\section{Concluding remarks}

An algorithm for geodesic motion on the constant-potential-energy hypersurface has been developed (Eq. (\ref{tt})). Single-precision simulations show that this algorithm, in conjunction with compensation for center-of-mass drift, is absolutely stable in the sense that potential energy, step length, and center-of-mass position are conserved for indefinitely long runs. The algorithm reproduces the $NVE$ radial distribution function of the LJ liquid, strongly indicating that correct configuration-space averages are arrived at in $NVU$ dynamics. 

Although $NVU$ dynamics has no kinetic energy providing a heat bath, it does allow for a realistic description of processes that are unlikely because they are thermally activated with energy barriers that are large compared to $k_BT$ (Paper II). In $NVU$ dynamics, whenever a molecular rearrangement requires excess energy to accumulate locally, this extra energy is provided by the surrounding configurational degrees of freedom. These provide a heat bath in much the same way as the kinetic energy provides a heat bath for standard Newtonian $NVE$ dynamics. 

The companion Paper (II)  compares the dynamics of the Kob-Andersen binary Lennard-Jones liquid simulated by the $NVU$ algorithm and four other algorithms ($NVE$, $NVT$, diffusion on $\Omega$, Monte Carlo dynamics), concluding that results are equivalent for the slow degrees of freedom. Paper II further argues from simulations and non-rigorous argumens that $NVU$ dynamics becomes equivalent to $NVE$ dynamics as $N\rightarrow\infty$.

\acknowledgments 
Useful input from Nick Bailey is gratefully acknowledged. The centre for viscous liquid dynamics ``Glass and Time'' is sponsored by the Danish National Research Foundation (DNRF).

\appendix

\section{Proof that the basic $NVU$ algorithm is symplectic}

This Appendix proves that the basic $NVU$ algorithm conserves the configuration-space volume element on the hypersurface $\Omega$ in the same sense as the $NVE$ algorithm conserves the configuration-space volume element. We view the basic $NVU$ algorithm  (Eq.  (\ref{nvu1})),

\begin{equation}
  \bR_{i+1}  = 2\bR_{i} - \bR_{i-1} -
   \frac{2\bF_{i} \cdot (\bR_{i} -
    \bR_{i-1})}{\bF_{i}^{2}}  \bF_{i}\,,
\end{equation}
as a mapping of $R^{6N}$ into itself. In the $6N$ dimensional configuration space of subsequent time-step pairs $\bS_i\equiv\{\bR_{i}, \bR_{i-1}\}$, the $NVU$ algorithm is

\begin{eqnarray}
  \textbf{S}_{i} \to \textbf{S}_{i+1} = \{\bR_{i+1}, \bR_{i}\} = \{ 2\bR_{i} - \bR_{i-1} -
   \frac{2\bF_{i} \cdot (\bR_{i} -
    \bR_{i-1})}{\bF_{i}^{2}} \bF_{i}, \bR_{i}\}.
\end{eqnarray}
The Jacobian of this map $\textbf{J}(\textbf{S}_{i} \to \textbf{S}_{i+1})$ is given by

\begin{equation} 
  |\mathbf{J}| = 
  \left| 
    \begin{array}{c c c c c c c c} \\
      & 2 - 2\frac{\partial \frac{\bF_{i}  (\bR_{i} - \bR_{i-1})}{\bF_{i}^{2}} F_{x_{1},i}}{\partial x_{1,i}} & 
      - 2\frac{\partial \frac{\bF_{i} (\bR_{i} - \bR_{i-1}) }{\bF_{i}^{2}}  F_{x_{1},i}}{\partial x_{2,i}}  & 
      ... & 
      -1 + 2\frac{\partial \frac{\bF_{i}\bR_{i-1}}{\bF_{i}^{2}}F_{x_{1},i} }{\partial x_{1,i-1}} & 
      2\frac{\partial \frac{\bF_{i}\bR_{i-1}}{\bF_{i}^{2}} F_{x_{1},i} }{\partial x_{2,i-1}}   & 
      ... & \\ \\
      & - 2\frac{\partial \frac{\bF_{i} (\bR_{i} - \bR_{i-1})}{\bF_{i}^{2}} F_{x_{2},i}}{\partial x_{1,i}}  & 
      2 - 2\frac{\partial \frac{\bF_{i} (\bR_{i} - \bR_{i-1})}{\bF_{i}^{2}} F_{x_{2},i}}{\partial x_{2,i}}  & 
     ... & 
      2\frac{\partial \frac{\bF_{i}\bR_{i-1}}{\bF_{i}^{2}} F_{x_{2},i} }{\partial x_{1,i-1}} & 
      -1 + 2\frac{\partial \frac{\bF_{i}\bR_{i-1}}{\bF_{i}^{2}} F_{x_{2},i} }{\partial x_{2,i-1}} & ... & \\ \\
      & \vdots & \vdots & & \vdots & \vdots & \\ \\
      & 1 & 0 & ... & 0 & 0 & ... & \\ \\
      & 0 & 1 & ... & 0 & 0 & ... & \\ \\
      & \vdots & \vdots & & \vdots & \vdots & \\ \\
    \end{array} 
  \right|.
\end{equation}
This may be regarded as a two-by-two block matrix consiting of blocks $\textbf{A},\textbf{B},\textbf{C},\textbf{D}$. The determinant of this block matrix is $|\textbf{J}| =|\textbf{A}\textbf{D} - \textbf{B}\textbf{C}| =|-\textbf{B}\textbf{C}| = (-1)^{M}|\textbf{B}|$, giving (where the index $i$ is dropped for brevity)

\begin{equation}\label{trick}
  |\mathbf{J}| = 
  (-1)^{M}\left| 
    \begin{array}{c c c c c c c} \\
      & -1 + 2\frac{F^{2}_{x_{1}}}{\bF^{2}} &  2\frac{F_{x_{2}}F_{x_{1}}}{\bF^{2}}  &  2\frac{F_{x_{3}}F_{x_{1}}}{\bF^{2}}  
      &  2\frac{F_{x_{4}}F_{x_{1}}}{\bF^{2}} & ... & \\ \\
      & 2\frac{F_{x_{1}}F_{x_{2}}}{\bF^{2}}  &  -1 + 2\frac{F^{2}_{x_{2}}}{\bF^{2}}  &  2\frac{F_{x_{3}}F_{x_{2}}}{\bF^{2}}  
      &  2\frac{F_{x_{4}}F_{x_{2}}}{\bF^{2}} & ... & \\ \\
      &  2\frac{F_{x_{1}}F_{x_{3}}}{\bF^{2}} &  2\frac{F_{x_{2}}F_{x_{3}}}{\bF^{2}}  &  -1 + 2\frac{F^{2}_{x_{3}}}{\bF^{2}} 
      &  2\frac{F_{x_{4}}F_{x_{3}}}{\bF^{2}} & ... & \\ \\
      &  2\frac{F_{x_{1}}F_{x_{4}}}{\bF^{2}} &  2\frac{F_{x_{2}}F_{x_{4}}}{\bF^{2}}  &  2\frac{F_{x_{3}}F_{x_{4}}}{\bF^{2}}  
      &  -1 + 2\frac{F^{2}_{x_{4}}}{\bF^{2}} & ... & \\ \\
      & \vdots & \vdots & \vdots & \vdots & \\ \\
    \end{array} 
  \right| = (-1)^{M}(\pm 1).
\end{equation}
Defining the unit-length vector ${\bf n} = \bf/\bF^{2}$, the last equality of Eq. (\ref{trick}) follows from 
$\textbf{B} = -\textbf{1} +2\textbf{n} \cdot \textbf{n}^{T} \Rightarrow \textbf{B}^{2} =  \textbf{1} +4\textbf{n} \cdot \textbf{n}^{T} - 4\textbf{n} \cdot \textbf{n}^{T} =\textbf{1}$. Since $|B|^2=|B^2|=1$, one has $| B| =\pm 1$. Thus the volume element transforms as

\begin{equation}
  d\bR_{i}d\bR_{i-1} = d\bR_{i+1} d\bR_{i}\,.
\end{equation}
This means that the basic $NVU$ algorithm conserves the volume element in the $6N$ dimensional configuration space, i.e., that the algorithm is symplectic just as the $NVE$ algorithm is.

\end{document}